# Estimating Network Effects Using Naturally Occurring Peer Notification Queue Counterfactuals

Craig Tutterow
LinkedIn Corp.
ctutterow@linkedin.com

Guillaume Saint-Jacques
LinkedIn Corp.
gsaintjacques@linkedin.com

**ABSTRACT**
Randomized experiments, or A/B tests are used to estimate the causal impact of a feature on the behavior of users by creating two parallel universes in which members are simultaneously assigned to treatment and control. However, in social network settings, members interact, such that the impact of a feature is not always contained within the treatment group. Researchers have developed a number of experimental designs to estimate network effects in social settings. However, it is not always feasible or desirable to run a network experiment due to engineering costs and/or ethical considerations. Naturally occurring exogenous variation, or 'natural experiments,' allow researchers to recover causal estimates of peer effects from observational data in the absence of experimental manipulation. Natural experiments trade off the engineering costs and some of the ethical concerns associated with network randomization with the search costs of finding situations with natural exogenous variation. To mitigate the search costs associated with discovering natural counterfactuals, we identify a common engineering requirement used to scale massive online systems, in which natural exogenous variation is likely to exist: notification queueing. We identify two natural experiments on the LinkedIn platform based on the order of notification queues to estimate the causal impact of a received message on the engagement of a message recipient (work anniversary and birthday announcements). We show that receiving a message from another member significantly increases a member's engagement, but that some popular observational specifications, such as fixed-effects estimators, overestimate this effect by as much as 2.7x. We then apply the estimated network effect coefficients to a large body of past experiments to quantify the extent to which it changes our interpretation of experimental results. The study points to the benefits of using messaging queues to discover naturally occurring counterfactuals for the estimation of causal effects without experimenter intervention. It also implies a potential benefit of involving data scientists in the system design process to maximize the informational benefits of software systems.

## 1 INTRODUCTION

When developing new products or features, companies, consumers and other stakeholders generally require evidence quantifying its effectiveness. The gold-standard for producing causal estimates is a randomized experiment, or A/B test. In randomized experiments, a random sample of the population is assigned to receive the feature (treatment group), whereas another random sample receives the status quo (control group). Randomized assignment is desirable because it allows the experimenter to produce a simultaneous counterfactual universe. Assuming the independence of treated units, we can then compare the two groups to estimate the effectiveness of the treatment in producing beneficial outcomes. However, in social network environments, people are capable of observing and interacting with others, such that a feature can have an indirect impact on those who do not receive it directly.

In these environments, researchers are unable to rely on estimates produced by standard statistical tests due to violations of the stable uniform treatment value assumption (SUTVA), which states that each user's behavior must be affected only by the treatment and not the treatment of others (Saveski et al. 2017).

A standard t-test in a situation with positive network externalities will result in downward biased coefficients. Consider, for example, a change in the feed interface that includes an oversized promotion in the feed encouraging the user to be the first to like their friend's post. This might have a zero or negative effect on the treated user's engagement. However, since the feature is likely to increase feedback received by the treated user's connection, it might have a positive second-order impact on engagement. Alternatively, situations with negative network externalities would result in upward biased coefficients. Consider the example of a chatbot. Giving a user access to an engaging chatbot may increase their engagement. However, if the user substitutes away from messaging their connections, this could produce a negative network externality, reducing messages received by other members, and leading to losses in downstream engagement.

A number of network-based randomizations, such as cluster-based assignment and peer encouragement designs have been developed in recent years to deal with this problem (Ugander et al. 2013, Eckles et al. 2016). However, network randomization is sometimes not feasible, e.g. in situations where the engineering investment required to randomize at the cluster level outweighs the expected benefits of incremental information. In other situations, network randomization might not be desirable, e.g. in situations that would harm user experience or fail to meet the 'no-greater-than-minimal-risk' (i.e. everyday risk) standard (Fiske & Hauser 2014).

Alternatively, observational estimation of peer effects, which does not require experimenter randomization, is made difficult in networked environments due to a number of time-varying, correlated processes, including: selective tie formation between similar alters (homophily), common exposure unobserved and correlated external events (confounding), and simultaneous influence of connected individuals (herding), or independent and simultaneous decision-making (the reflection problem) (Manski 2000, Hartmann et al. 2008, Shalizi & Thomas 2011, Aral & Walker 2012).

In this paper, we demonstrate an approach using natural exogenous variation and an instrumental variable estimator to overcome the dual constraints of a) the SUTVA violation in social network environments, and b) absence of experimenter randomized assignment. The approach described here, subject to a number of assumptions, allows us to recover causal estimates of peer effects from historical observational data. We show that common observational model specifications (OLS, fixed-effects) overestimate the second-order impact of the feature by a factor of 2-4x in the absence of high dimensional controls (Eckles & Bakshy 2017). On the other hand, a naturally occurring 'encouragement design,' whereby a user's (ego) connections are encouraged to reach out to ego via LinkedIn messaging, in combination with natural randomization in the notification queue allows us to generate more accurate causal coefficients as to the impact of receiving a message on the focal user's engagement.

We backtest our estimates with past experimental data to quantify the extent to which messaging network effects impact our interpretation of results ex-post. We find that engagement estimates are substantially impacted by incorporating the estimated network effect coefficients. Of top 100 most impactful messaging experiments, 49 also produced statistically significant changes in our measure of engagement (pageviews). Of those experiments, we found that the standard A/B test results over or underestimated the net engagement effect by an average of approximately 16%. Accurately estimating network effects is essential in social network environments, especially in experiments that produce large network externalities (more or less messages, likes, comments, or other types of observable and interactive social behavior). This result also illustrates the potential importance of incorporating network effect estimators directly into standard A/B test reporting platforms in order to better educate product teams.

In section 2, we describe related work on this topic. Section 3 proceeds to describe the methodological approach and the empirical setting (messaging on the LinkedIn platform). Section 4 presents a comparison of results from the instrumental variable estimator in the peer encouragement natural experiment, as compared with other observational techniques. Finally, we conclude with a discussion of potential extensions of this line of work.

## 2  RELATED WORK

Peer encouragement experimental designs, described in Eckles et al. (2016) typically consist of experimentally 'nudging' an individual's connections to take a certain action in order to estimate the peer behavior impact on the original individual. Prior studies in social psychology measured peer effects using 'confederates' who exert influence on the participant according to instructions from the experimenter in an artificial laboratory environment (e.g. Asch's 1956 conformity experiment). Web-based social platforms make peer-effect experiments feasible in more natural settings and at increased scale. For example, in Eckles et al. (2016), the peers of certain individuals see a modified feed interface, which incentivize them to provide more feedback to ego, in the form of more likes or more comments. This is then used to causally identify the ego's reaction to receiving more feedback.

Such experiments, while ideal because of randomized assignment, may be quite costly: they require the ability to treat members at the edge level in the context of a specific feature, which may require implementation at both the level of the experimentation platform as well as at the level of the product. In other words, the platform has to show content originating from different egos with a different interface.

Alternatively, cluster-based randomization (Ugander et al. 2017) requires costly engineering efforts to dynamically assign treatments at the network level, and significantly reduces the number of observations available to researchers. While variance reduction techniques have been developed to improve statistical power (Ugander et al. 2017), this approach is often too expensive to implement unless one expects large effect sizes.

Observational estimation of peer effects, on the other hand, is notoriously difficult (Manski 2000, Aral et al. 2009, Shalizi & Thomas 2011, Aral & Walker 2012). Multiple identification strategies have been employed to deal with the issues of endogenous tie formation, correlated unobservables, and simultaneity that confound most peer effect studies (Nam et al. 2010, Nitzan & Libai 2011, Bollinger & Gillingham 2012, Yoganarisimhan 2012, Phan & Airoldi 2015). In the marketing domain, this is required to obtain more precise estimates of the social spillover and viral multiplier effects of advertising, thereby allowing advertisers to come to a more accurate estimation of return on investment (Nair et al. 2010, Nam et al. 2010). However, these approaches usually require highly context and dataset specific modeling adjustments and assumptions. Furthermore, prior work has shown that, in the absence of high-dimensional controls, fixed-effects and other observational estimators still run the risk of substantially overestimating the impact of peer behavior (Eckles & Bakshy 2017).

Our approach relies on a similar causal path to randomized peer assignment: changes affecting an individual's connections, which in turn affect the individual. The main difference is that we rely on the inherent randomness of notification timing, bypassing the need to organize experiments or require special investment at the level of the product.

This is helpful because many peer effect experiments may run the risk of creating a negative user experience: for example, delaying or intercepting actual messages between users would not be acceptable. On the other hand, leveraging randomness that is inherent to product and engineering design does not add such a risk. A parallel of this risk can be seen in the discussion around the Facebook emotional contagion study (Kramer et al. 2014), which produced valuable findings on how emotions spread through a social network, but has been criticized in the popular press. A separate team drew on a natural experiment, using rainfall as an instrumental variable to arrive at a similar conclusion using only observational data (Coviello et al. 2014).

Finally, we attempt to build on the ad-hoc and contingent modeling approaches employed in most observational studies, by exploring how notification queues and other computational systems that split up workloads into parallelized 'batches' often contain 'natural' exogenous variation that can be used to recover causal estimates using observational data. We hope that this will encourage more thinking as to how computational systems or methodologies could be designed to further reduce the search costs associated with natural experiments (Sharma et al. 2016).

## 3 PEER ENCOURAGEMENT NATURAL EXPERIMENT AND INSTRUMENTAL VARIABLE REGRESSION

### 3.1 Data and Setting

The LinkedIn mobile and web applications contain a messaging feature that allows users to send messages to each other. In order to facilitate conversations between users, LinkedIn occasionally sends notifications about member milestones such as new jobs, promotions, job anniversaries, and birthdays.

The scale of notification requests flowing through large social networking platforms precludes simultaneous delivery, and therefore requires queueing systems to manage throughput. For instance, LinkedIn's member notification gateway, ATC, processes over one billion requests per day (Shi & Fuad 2018). The fact that some notifications are sent earlier than others produces valid counterfactuals for estimating the causal impact of certain peer behaviors without the need for randomized assignment insofar as a) the observation window is isolated as a subsegment of the overall delivery time window, and b) the notification ordering is not correlated with the outcome of interest.

In the absence of exogenous variation, an experimenter would need to construct an experimentation platform that randomizes peers to receive the notification or not. Figure 1, below, shows the design of a randomized peer encouragement experiment. Those egos assigned by the experimentation platform to treatment would have their peers (red nodes) receive the notification encouraging them to send a message. This encouragement would be withheld from those randomly assigned to the control group. An analyst would then estimate differences in behavior between the two ego nodes over the experiment window.

Figure 1: Randomized peer encouragement experiment design

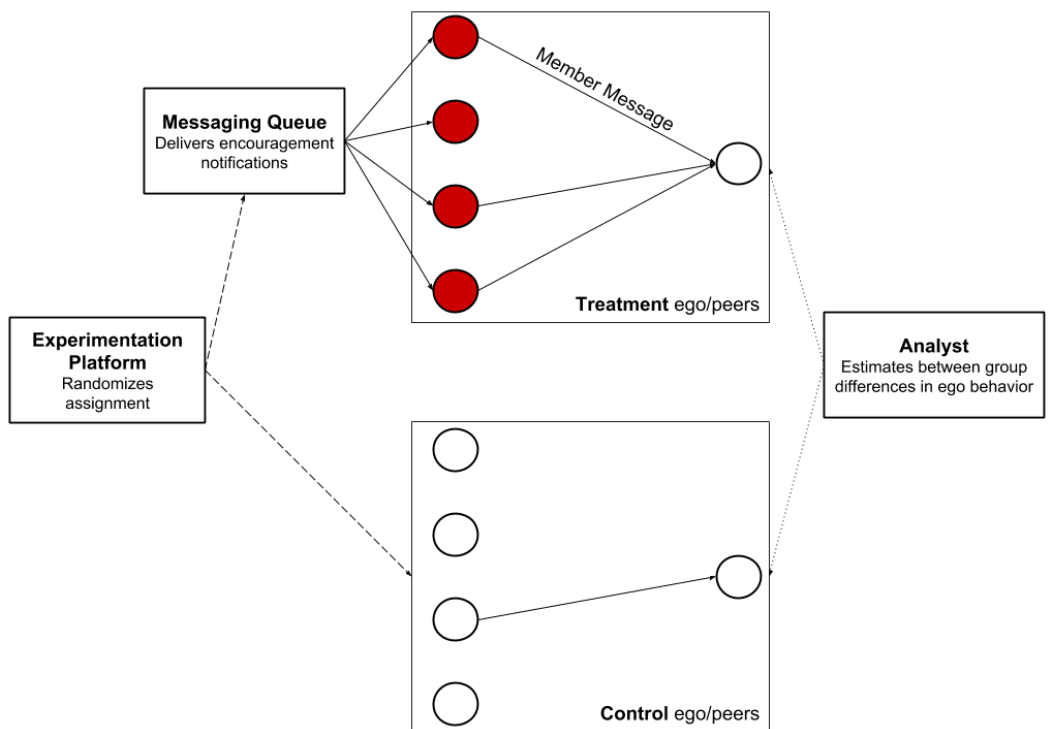

However, in the presence of a notification queue that is uncorrelated with the dependent variable, we have the potential to observe a valid counterfactual in the absence of experimenter randomization. The 'natural experiment' setup requires comparing an early notification group to a later one over a shortened observation window (see figure 2 below). However, notifications are ultimately delivered to all members over the intended time period, according to engineering specifications. That is, treatment is not withheld from any member. This provides the benefits of causal inference without the implementation costs of network randomization or ethical concerns surrounding the withholding of a feature.

Figure 2: Notification queue based observational peer encouragement design.

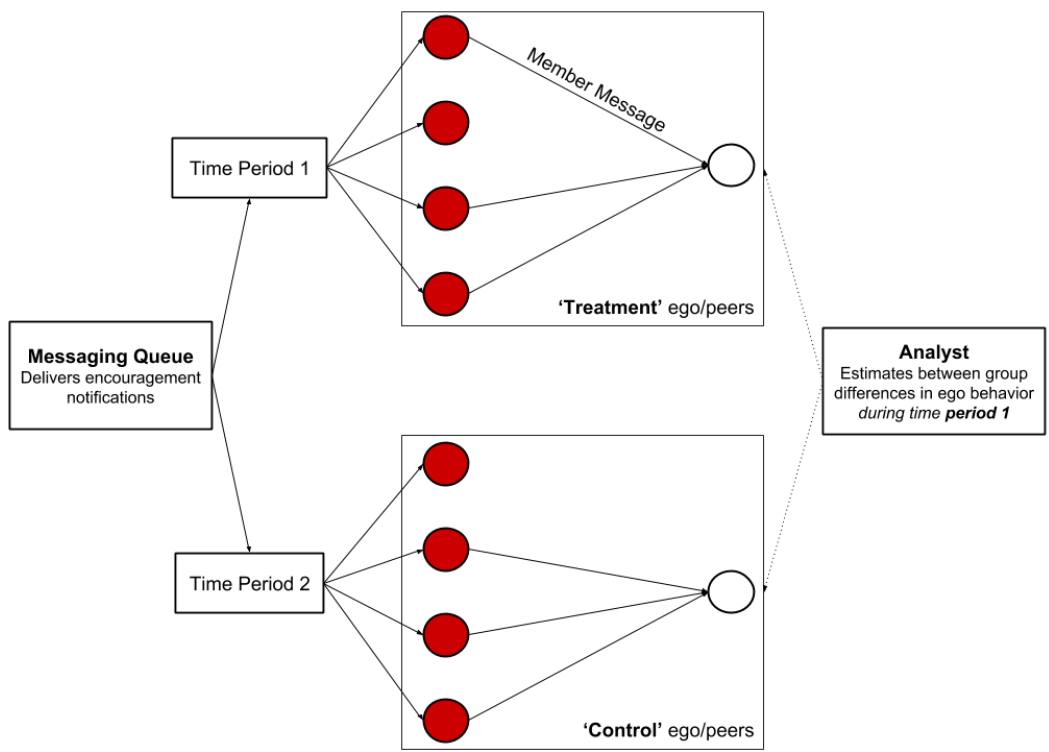

Interestingly, the LinkedIn context offers two natural experiments that can be used to generate causal estimate of the impact of receiving a message on user engagement. First, when entering work experience and current positions on their profile, users are able to enter the month and year in which they started the position (as is customary on resumes). LinkedIn derives work anniversary information from member profiles at a monthly level. An offline job evenly schedules message load over the days in that month, subject to other optimization and load-balancing techniques. When scheduled, all of a member's connections who are opted in will receive a notification suggesting: 'Congratulate [connection name] for 2 years at [employer]' with a clickable prompt reading 'Say congrats,' which autopopulates a messaging draft (see figure 3 below).

Since the specific week that all of a member's peers are notified about their work anniversary is randomly distributed throughout the month, we are able to compare the behavior of users who had the notification go out one week to those who had it go out in another. In particular, we compare messages received and engagement (measured in pageviews) for members over the second and third weeks of January 2018 (observation period) for those who had the notification go out to their contacts in the second week of January (treatment group) versus those who had the notification go out in the fourth week of January (the control group, who had the notification sent outside of the observation window).

We choose engagement as our dependent variable due to 1) its salience for engagement driven social network businesses, and 2) the fact that it is adjacent to, rather than directly related to the social feature being studied (messaging). Many studies of spillover effects restrict themselves to the interaction being measured (e.g. the impact of messages received on messages sent). The largest spillover effects will be found in metrics related to a particular mode of interaction. We wish to show that the approach can be generalized to metrics outside of the type of interaction itself. We choose pageviews as our specific measure of engagement due to its interpretability and the simplicity of its definition. However, we ran tests on engagement metrics with more complex definitions and found the results to be robust over all of them.

We chose to start the observation period in the second week of January rather than the first in order to minimize any holiday effect from the New Year's period. We removed outliers (top 1% of pageviews) to reduce bias towards extreme values. We also performed an A/A test, comparing pageviews in the week prior to the start of the observation window to confirm that the queue order was not associated with the outcome of interest (t = 1.15, p = 0.25) or a session-based measure of engagement (t = 0.92, p = 0.36), with average engagement being slightly higher in the control group than the treatment.

Figure 3: Work anniversary messaging encouragement.

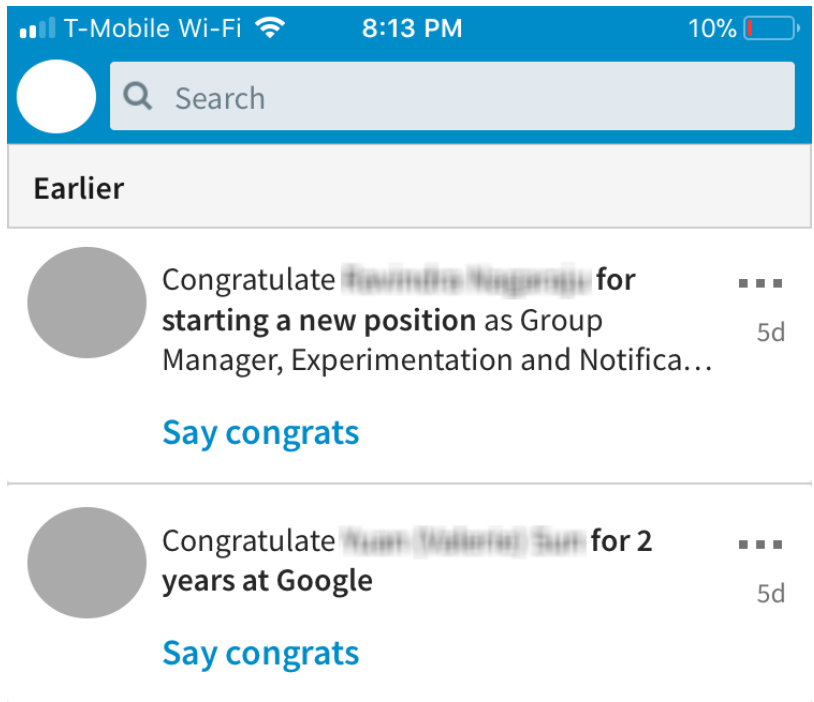

Figure 4: Notification schedule and observation window for instrumental variable regression.

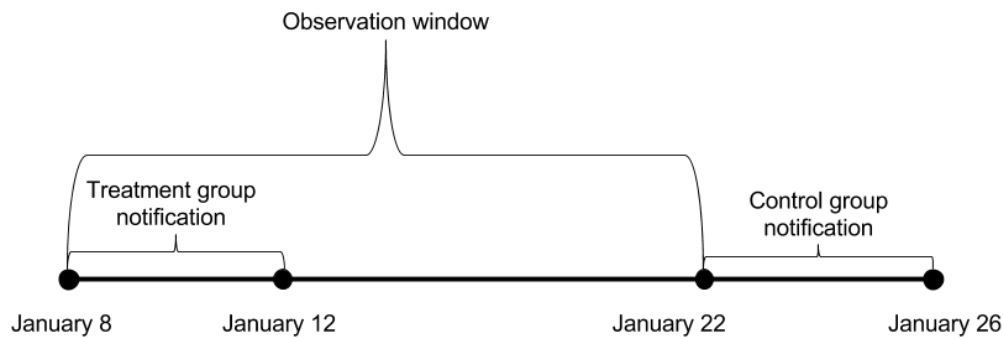

Since the week the notification goes out is arbitrarily determined and these two groups are comparable on all other dimensions, we can use this 'naturally occurring' (i.e. non-experimenter induced) exogenous variation in combination with an instrumental variable model to estimate the causal impact of a received message on a user's engagement.

A second source of natural variation can be found in birthday notifications. Birthdays are commonly used as instruments in the economic literature (Angrist 1992, Angrist & Krueger 2001). While birth dates have been shown to systematically vary at the annual level (across months), where the birth date falls within smaller time scales does not have a systematic relationship with most conceivable outcomes. Birthday notifications differ from work anniversary ones because the user provides their specific date of birth, so the day in which the notification goes out to peers is pre-determined. However, date of birth is typically considered a valid instrument because parents (and their children) typically have limited control over the birth date on smaller time scales. Week of birth here serves as an instrument that effectively 'randomizes' whether messaging encouragement notifications will go out to a user's peers in the observation period or not. Despite being similar in all other respects, those born in the second week of January are therefore more likely to receive messages over the following two weeks, whereas those born in the fourth week of January are not. Because of this, we are able to compare the engagement of those who had birthday notifications sent to their connections during the second vs. the fourth week of January in much the same way as we did with work anniversary notifications. Figure 4, below, shows the birthday messaging notification.

Figure 5: Birthday messaging encouragement.

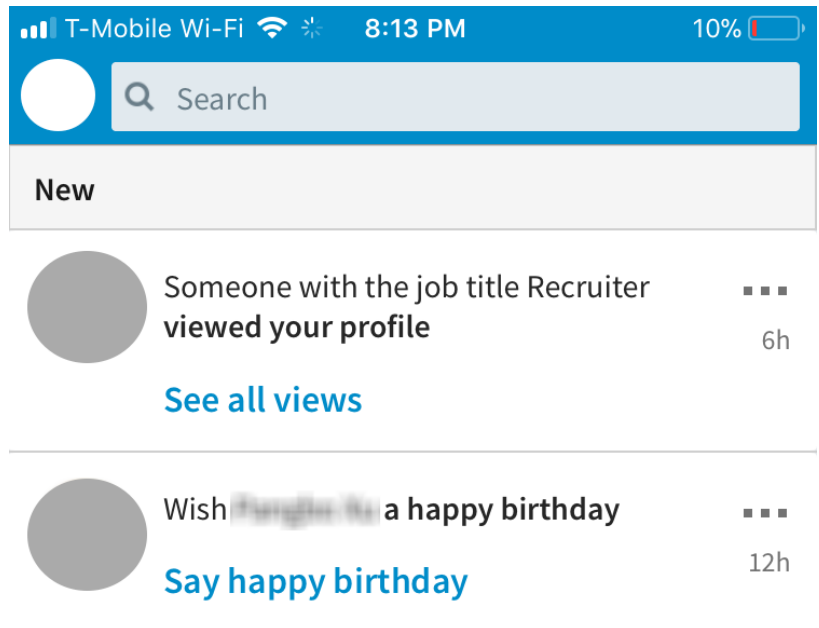

## 3.2 Two-Stage Least Squares Instrumental Variable Regression

Instrumental variable regression is a common method in econometrics for estimating the causal impact of an independent variable on an outcome of interest in the event of a natural experiment. Natural experiments are defined by Angrist & Krueger as 'situations where the forces of nature or government policy have conspired to produce an environment somewhat akin to a randomized experiment' (2001). A requirement for using instrumental variable estimator is the possession of an 'instrument' which is a variable that proxies for random assignment to the treatment of interest. The instrument (a peer encouragement notification) must be causally associated with the independent variable of interest (e.g. the number of messages received by ego), but not directly causally related to the outcome of interest (ego's engagement).

In the LinkedIn context, when a member has a work anniversary or a birthday, the platform generates a notification to her connections, suggesting congratulating the member on the occasion. We use the timing of this notification as an instrument for the number of messages received by a member. The messaging 'nudge' sent to a node's connections in this context is an appropriate instrument because it is both a) randomly assigned to a given week within the month, and b) unrelated to ego's engagement (as ego does not see or receive the notification itself, only the messages it induces from peers).

In seeking to identify the causal effect of receiving a message on user engagement, we argue that the timing constitutes a valid instrument for the number of messages received by a member, because it satisfies the exclusion restriction (that the variable is only causally related to the independent variable of interest, and not the dependent variable) and, in our specification, has a strong first stage (or causal association with the number of messages received by ego).

Two-stage least squares regression uses this exogenous variation to 1) estimate the impact of the treatment on the number of messages an ego node receives from their peers, and 2) estimate the indirect impact messages received as a consequence of the instrument on ego's engagement (as measured by pageviews). In a simple specification with no controls and constant treatment effects, stage 1 is estimated from the following equation:

$$RM_i = \gamma T_i + \varepsilon_i$$

where $M_i$ is the number of messages received by individual $i$ and $T_i$ is a dichotomous variable that captures whether a notification was sent out to her peers during the period of interest (i.e. an indicator for natural assignment to treatment or control group). $\gamma$ therefore captures the causal effect of the notification on number of messages received. We then estimate the second stage, substituting the number of messages received by a member by its value as predicted by the first stage. This is to identify the effect only based on the variation in messages received that was caused by the treatment, and not its "natural" variation, which might be correlated and leads to bias in a regular OLS specification. Stage 2 can be written as:

$$Pageviews_i = \beta \widehat{M_i} + \varepsilon'_i$$

where $\beta$ captures the causal effect of receiving one additional message on page views.

*3.2.1 Exclusion restriction (1-random assignment)*

Our identification strategy relies on using the fact "work anniversary" as well as "birthday" notifications are sent at a more or less random time of the month to members' connections. This means that, within the group of all members who have an anniversary during January, we can define a dummy variable "anniversary notification sent during the second week of the month", and that this dummy variable is independent of all other covariates and potential confounders.

*3.2.2. Exclusion restriction (2-no other effects of the instruments)*

For the exclusion restriction to be satisfied, it is also important to make sure that there is no "causal back-channel", i.e. that the instrument only impacts our endogenous variable (the number of messages received by a member). This is built into the design of the notification: the only action recommended is "send a message". There is no recommendation to post, or comment, or interact with the member in any other way. Of course, it is conceivable that some minute violations might happen, for example if upon seeing the notification, a member decides to pick up the phone to congratulate her connection in person, and the receiver visits LinkedIn as a result. Given the small likelihood and effect size of such behaviors, we argue that the exclusion restriction in this case is a reasonable assumption. As Eckles et al. (2016) note, the exclusion restriction is particularly plausible in peer encouragement designs, as ego is not exposed to the treatment itself.

*3.2.3. First Stage*
Instrumental variable regression requires instruments that have a sufficiently strong impact on the independent variable of interest. Weak instruments do not yield accurate coefficients. Out of a concern for the possible weakness of the instrument, we make sure that we have a strong first stage relationship between the encouragement notification to peers and messages received by ego. These notifications are a very strong driver of messaging, and account for a significant portion of total messages received in any given week. We also provide the results of a few weak instruments tests below, all of which strongly reject the null hypothesis that messaging encouragements constitute a weak instrument.

Our identification strategy and assumptions can be summarized with the following DAG. In the first stage, we estimate the impact of peer notifications on the number of messages received by ego. Second, we estimate the indirect impact of messages received on ego's engagement (pageviews). Finally, we assume that notifications sent to peers (and therefore, not directly seen by ego) have no direct impact on ego's engagement, and that pageviews during the period have no association with the distribution of peer notifications.

Figure 6. Graphical Model of Peer Encouragement Instrumental Variable Model

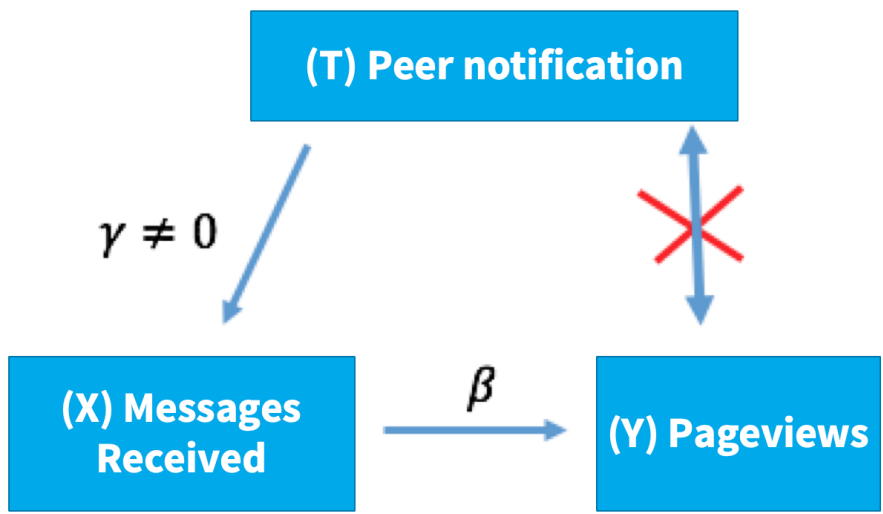

To the extent possible, we verify this empirically by comparing the pageviews of those in our 'treatment' and 'control' groups, where we observe no difference in engagement in the week prior to the observation period using engagement measures for both sessions (t = 0.92, p = 0.36) and pageviews (t = 1.15, p = 0.25), with the average amount of engagement being slightly higher in the control group compared to the treatment group.

In general, it seems that many notification systems, which include a component of randomness in their timing, are good candidates to be leveraged in an instrumental variable framework, and can provide "free" sources of random variation, without the need to organize an experiment.

The following section introduces the fixed effects estimator, a common technique to correct for omitted variable bias in observational studies. We then discuss the costs and benefits of the considered approaches, and compare the coefficients produced by the two observational estimators: instrumental variable and fixed-effects regression.

### 3.3 Fixed-Effects Regression

Fixed-effects regression is a common approach for reducing bias by controlling for time-invariant person-specific confounders. The fixed-effects estimator is desirable for modeling within-person change due to the fact that the respondent-specific intercept absorbs all person-specific, time-invariant confounders, such as work experience, stable personality traits, educational background, etc. that might be associated with changes in engagement. By including a person-specific dummy variable, we effectively reduce the potential for omitted variable bias to only time-varying confounders. We also include time-specific fixed effects for week, which controls for period-effects that might have led all users to engage at different levels based on what was going on at that time. The fixed-effects model can be written as:

$$Pageviews_{it} = \beta_{messages\ received_{it}} + \sigma_i + \tau_t + \varepsilon_{it}$$

where $Pageviews_{it}$ is the dependent variable for user i at time t, $\beta_{messages\ received_{it}}$ is the number of messages received by ego i from their peers at time t, $\sigma_i$ is the user-specific fixed-effect, and $\tau_t$ is the time-specific fixed-effect.

Figure 5 depicts the observation window for the fixed effects regressions, which were performed on data from December to avoid overlap with the treatment encouragements.

Figure 7: Fixed-effects observation windows.

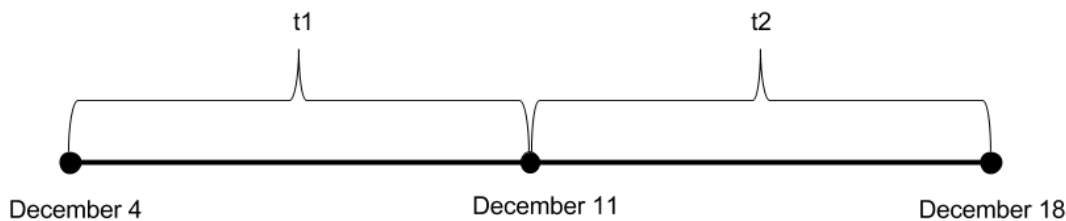

All users who had either a birthday or work anniversary notification go out in the December observation window were removed from the sample.

### 3.4 Costs and Benefits of Different Approaches

Table 1, below, compares the benefits and costs of fixed-effects regressions, instrumental variable regression and the directed edge-level randomization used in peer encouragement designs.

Fixed effects are the most general purpose method but tend to produce upwards biased estimates if high-dimensional controls are not used (Eckles et al. 2017). Natural experiments are a less invasive method for getting an unbiased estimate of causal treatment effect, but there are significant search costs associated with finding empirical settings with naturally occurring exogenous variation. Edge-level randomization, as used in other peer encouragement designs are the gold-standard for inference, but require significant infrastructure investments, and might not be desirable in instances where the treatment would lead to negative user experience (e.g. delaying messages).

Table 1: Benefits and Costs of Different Methods for Estimating Network Effects

| Approach | Benefits | Costs |
| --- | --- | --- |
| Fixed-Effects Regression | General purpose, low cost, can be applied to any panel dataset | Peer effect estimates typically upward biased unless able to include high-dimensional controls |
| Natural experiment (Instrumental Variable Regression) | Generates accurate causal estimates of peer effects in the absence of experimenter randomization. | Hard to find natural exogenous variation. Unlikely to have random experiment for every research question. |
| Edge-level randomization | Gold standard for experimental measurement of peer effects. Easy to scale to many experiments once infrastructure is established. | High engineering fixed costs to setup. Can result in temporarily negative user experience. |

### 4. RESULTS FROM MESSAGING ENCOURAGEMENT NATURAL EXPERIMENT

This section uses four models to estimate the impact of receiving a message on a focal user's engagement. To measure engagement, we take an intuitive metric – pageviews – as our dependent variable. A pageview is generated each time a user visits LinkedIn or clicks a navigation button or link that takes them to another page on the site. All data used for the study was deidentified. After analysis was complete, the dependent variable (pageviews) was multiplied by an arbitrary constant to obfuscate the coefficient values for external reporting.

The four models under comparison are:
- *OLS*: A univariate ordinary least squares regression model, with messages received as the only predictor.
- *OLS with controls*: A multivariate OLS model with control variables for connections decile, country, industry and years since entering the workforce.
- *User Fixed-Effects*: An OLS model with a time parameter and dichotomous variable for each member to control for all time-invariant confounders.
- *Instrumental Variable*: A two-stage least squares model that estimates the effect of messages received on user engagement using peer encouragement treatment status as the instrument

Table 2 compares the results from each of these models for the birthday messaging natural experiment.

In all models, we see a strong, positive relationship between messages received and pageviews. However, the univariate OLS model overestimates the relationship by a factor of 8.2 (14.02/1.72). Including the aforementioned control variables reduces this only slightly (to 7.4x above the IV estimate). On the other hand, the fixed effects model performs significantly better, but still overestimates the causal effect by a factor of 2.7x (4.68/1.72).

Table 2: Birthday Encouragement: Regression output of Member Pageviews on Messages Received

| | Dependent variable: Pageviews | | | |
|---|---|---|---|---|
| | OLS (12/4-12/11) | OLS (12/4-12/11) | User FE (12/4-12/17) | Instrumental Variable (1/8-1/22) |
| Messages Received | 14.025*** | 12.689*** | 4.679*** | 1.716*** |
| | (0.170) | (0.176) | (0.018) | (0.078) |
| Includes Controls | No | Yes | No | Yes |
| Includes User FE | No | No | Yes | No |
| Includes Time FE | No | No | Yes | No |
| $R^2$ | 0.152 | 0.193 | 0.912 | 0.202 |
| Adjusted $R^2$ | 0.152 | 0.185 | 0.824 | 0.202 |
| F-Statistic | 6778*** | 23.97*** | 10.39*** | 134.4*** |
| Note: | | | | *p<0.05; **p<0.01; ***p<0.001 |

Table 3, below, presents results comparing the results for the work anniversary natural experiment. Here we see a similar, but less pronounced pattern with OLS and fixed-effects models on observational data overestimating the effect of receiving a message on engagement by a factor of 1.7-2.7x.

Table 3: Work Anniversary Encouragement: Regression output of Member Pageviews on Messages Received

|  | Dependent variable: Pageviews | | | |
| --- | --- | --- | --- | --- |
|  | OLS (12/4-12/11) | OLS (12/4-12/11) | User FE (12/4-12/17) | Instrumental Variable (1/8-1/22) |
| Messages Received | 6.953*** | 6.379*** | 4.110*** | 2.484*** |
|  | (0.039) | (0.039) | (0.007) | (0.050) |
| Includes Controls | No | Yes | No | Yes |
| Includes User FE | No | No | Yes | No |
| Includes Time FE | No | No | Yes | No |
| $R^2$ | 0.128 | 0.183 | 0.910 | 0.136 |
| Adjusted $R^2$ | 0.128 | 0.181 | 0.819 | 0.136 |
| F-Statistic | 31180*** | 121.1*** | 10.07*** | 527*** |

Note: *p<0.05; **p<0.01; ***p<0.001

Table 4 presents the results from the first stage of the 2SLS procedure described above. This indicates a strong positive effect of peer assignment to treatment on messages sent from peers to ego during the observation window, alleviating concerns regarding the strength of the instrument.

Table 4: Stage 1 Regression on Number of Messages Received by Ego From Peers on Peer Encouragement Treatment Status

|  | Dependent variable: Messages Received by Ego | |
| --- | --- | --- |
|  | Birthday | Work anniversary |
| Peer treatment | 8.925*** | 4.397*** |
|  | (0.068) | (0.029) |
| $R^2$ | 0.185 | 0.0514 |

Note: *p<0.05; **p<0.01; ***p<0.001

Figure 6, also shown below, depicts the coefficients produced by each model in each natural experiment. Interestingly, the OLS models produce much higher estimates for the birthday notification than the anniversary one, but when fixed effects are introduced, the two seem to perform similarly. When we look at the IV specification, this association appears to be reversed, with the work anniversary messages generating more downstream pageviews than birthday ones.

Figure 8: Comparison of pageview coefficient for message received by model and encouragement type.

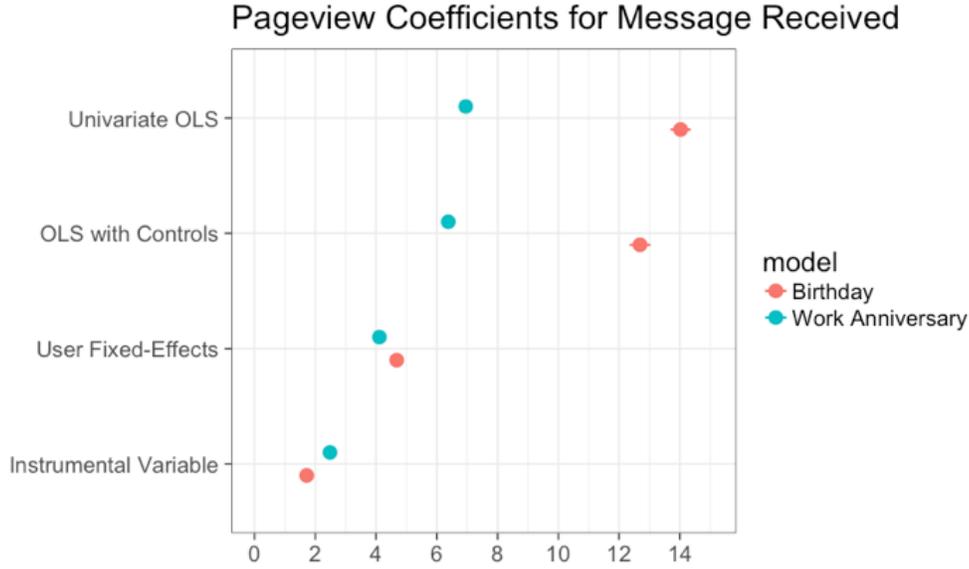

The difference between the estimates obtained with the birthday instrument and the ones obtained with the work anniversary instrument reveal a subtlety of instrumental variables. When the treatment effect is not constant (i.e. each individual has their own idiosyncratic reaction to being treated), the IV estimate returns a Local Average Treatment Effect (Angrist & Pischke 2009), which depends on the population that was affected by the instrument. The difference here suggests that individuals who are affected by birthday messages are different from the ones affected by work anniversary messages. This is likely due to the fact that birthday information is an optional and nonobvious profile field, used more often by power users who are already highly engaged with the site, and therefore may have less sensitivity to network effects. On the other hand, the work experience profile field is highly utilized, including by less engaged users, whose engagement might be more sensitive to network effects.

### 4.1 BACKTESTING ON PAST EXPERIMENTS

To estimate the impact of messaging network effects on our interpretation of past results, we backtested on experiments run on the LinkedIn experimental platform during the 2017 calendar year. The delta, or impact, of a given experiment, $i$, can be written as follows:

$$\Delta_{Pageviews_i} = \frac{(Pageviews_T - Pageviews_C)}{Pageviews_C}$$

where $Pageviews_T$ is the average number of pageviews of a user assigned to the treatment group and $Pageviews_C$ is the average number of pageviews of a user assigned to the control group. We apply the learned coefficient from Table 3 to messages sent (and therefore received by other members) to generate adjusted deltas based on a combination of main effect of the experiment, in addition to the predicted network externalities caused by the increase or decrease in messaging activity. The adjusted delta can be written as follows:

$$\Delta'_{Pageviews_i} = \frac{(Pageviews_T + \beta \Delta_{Messages_i} R - Pageviews_C)}{Pageviews_C}$$

where $\beta_{SS}$ is the coefficient representing the network effect of messages sent on a receiver's pageviews, and R represents a discounting factor to account for network interference. Since we assume that messages are sent proportionally to those in the treatment and control groups, $R = 1 - P(treatment)$. The intuition for the discounting parameter is as follows: the larger the proportion of members assigned to treatment, the more the network effect is already observed, as incremental messages are more likely to be sent to those also in treatment. $\Delta_{Messages_i}$ is the delta in messages sent for experiment i:

$$\Delta_{Messages_i} = \frac{(Messages\ Sent_T - Messages\ Sent_C)}{Messages\ Sent_C}$$

Therefore, the equation for $\Delta'_{Pageviews_i}$ gives us the adjusted experiment delta based on both the main effect of the experiment and the estimated first-order network effect on message receivers.

To establish the upper bound on the bias we might expect by failing to measure network effects, we looked at the top 100 experiments ranked based on the magnitude of their impact on messages sent by treated members. These 100 experiments lifted messages sent by an average of ±8.3%, and a median of ±4.6%. Of the 100 top experiments in 2017 moving messages sent, we found that 49 also significantly moved pageviews for the treatment group (p ≤ .05).

We use this group of experiments to compute the average error between the original delta (which did not take into consideration network effects), and the adjusted delta (which does) over the experiments included in our sample. This can be written as follows:

$$\varepsilon = \frac{\sum_{i=n}^{i=1} \frac{|\Delta_{Pageviews_i} - \Delta'_{Pageviews_i}|}{\Delta_{Pageviews_i}}}{n}$$

When we retrospectively apply the network effect coefficient from the instrumental variable model to adjust the deltas from these experiments, we find an average error of 15.9% (median = 13.4%). In other words, on average, when we look at the most impactful messaging experiments, the results of a standard A/B test will over or underestimate the impact of the feature on pageviews by approximately 15.9%.

The distribution of error rates can be seen in the below histogram (figure 7). We can see here that some experiments over or underestimated the effect size of the feature on engagement by as much as 50%. We also see a negative skew, which reflects the fact that most experiments increase messages sent and therefore underestimate their impact on engagement.

Figure 9: Histogram of Effect Size Error Rates After Network Effect Adjustments

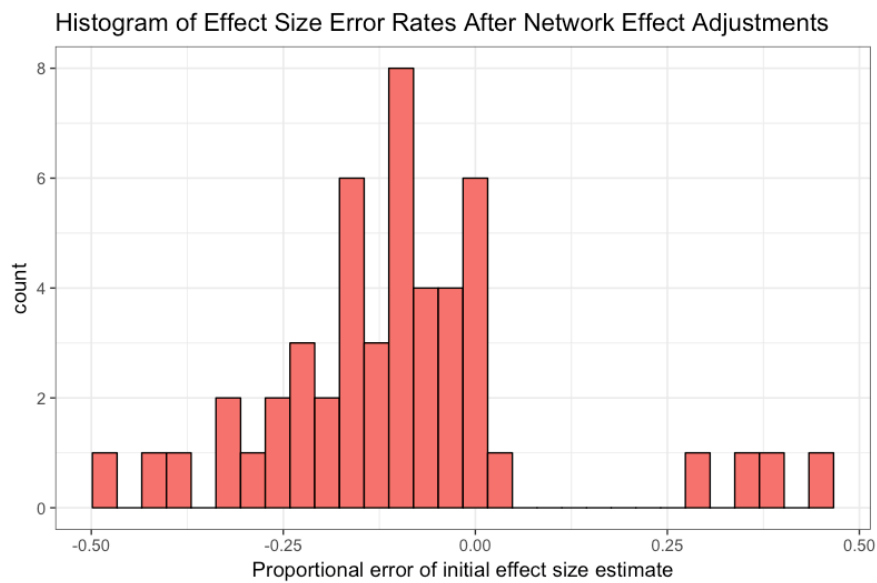

This is only the first-order network effect, and therefore does not take into account any compounding. Furthermore, this model only takes into account one type of viral action, and so is only a partial and incomplete view of a far-reaching ecosystem of potential network effects caused by social influence and observation in the LinkedIn setting (e.g. indirect impact from feed activity).

There are two noteworthy drawbacks of our analysis here. First – our research context required an observational design and therefore precluded a comparison against a 'ground-truth' dataset produced with a randomized experiment. Prior work has shown that the fixed-effects procedure used above overestimates peer effects in the absence of high-dimensional controls. In addition to the extensive econometric literature on instrumental variable regression (Angrist & Krueger 2001), prior methodological work (Shalizi & Thomas 2011) gives us confidence that fixed-effects results are overestimated, and that the instrumental-variable estimate is a better indicator of the causal effect of receiving a message on subsequent engagement. Second, because we needed to search for situations with naturally occurring counterfactuals, we are limited to two types of messaging encouragements (work anniversary and birthday). The generalizability of the estimates produced here is only valid insofar as these types of messages are representative of other types of messages. It is possible that these messages underestimate the broader impact of messages received on the platform as they are the result of a nudge rather than spontaneous interaction with the product. For this reason, spontaneous messages could potentially signal higher intent and relevance, and therefore have a larger network effect than encouraged ones.

## 5 CONCLUSION AND DISCUSSION

This study has explored the network impact of messages on downstream engagement by message receivers. Using a natural experiment, we were able to recover causal estimates using observational data. The estimated coefficient can then be combined with experimental data to generate predicted values as to the net engagement effect of a feature in a way that takes into consideration both the direct impact of the feature on the person who sees it, as well as its indirect impact on their peers. The validity of such predictions ultimately depend on how representative the instrument derived from the natural experiment is, and direct estimation with a natural or randomized experiment is required to generate a true causal estimate for a specific feature. Nonetheless, this provides useful information as to the extent to which network effects may alter our interpretation of experimental results, as well as our decisions around tradeoffs between engagement and virality. For example, one might decide to ramp a feature that has a slightly negative engagement effect on the treatment group, insofar as this is counteracted by a positive indirect engagement effect caused by the increase in messaging.

Increasing attention is being paid to the tradeoffs between the informational benefits and implementation costs of observational and experimental methods. In some cases, experimenter randomization is not feasible (due to engineering costs) or desirable (due to ethical concerns). Salganik (2017) suggests that 'minimally invasive' methods should be preferred to experimenter randomized assignment in situations where users face substantial risk and provides a principles-based framework for making decisions around these tradeoffs (based on counterfactual situations in which the research is conducted or not conducted using different methodologies). In situations with significant costs to member experience (e.g. involving the delivery of messages), it is appropriate to look for a natural experiment, which allows the researcher to observationally estimate causal network effects in the absence of experimenter assignment. In cases where no natural experiment is available, fixed effects or matching will reduce the upward bias of a simple OLS model. However, in the absence of high-dimensional controls (Eckles & Bakshy 2016), fixed-effects still overestimates the causal effect in our setting by a factor of two.

While this study focused on messages, the methods used here could be extended to estimate and compare the size of network effects across different types of viral actions (e.g. liking a post, commenting on a post, or creating a post). The results of our backtest suggest that other types of viral actions are likely to have even more substantial indirect impact on engagement, in part due to their one-to-many public nature. While one-to-many viral actions may affect individuals less, they also reach a much larger audience, which has the potential to produce a larger global effect.

Natural experiments trade off the engineering costs and ethical considerations associated with network randomization with the search costs of finding situations with natural exogenous variation. To mitigate the search costs associated with discovering natural counterfactuals, we identify a common engineering design used to scale massive online systems, in which natural variation is likely to exist: message queueing. While message queueing often occurs on very small, subsecond timescales, in some cases it is scheduled over a number of days or weeks (as is the case here). Even in situations where queueing and delivery occurs on smaller timescales, such natural variation could be useful when it has a significant impact on the independent variable of interest and is unassociated with the outcome. For instance, notification schedules are often based on time zones. As time-zone borders are often based on historical accidents, many cities with the same longitude and other attributes are located in different time zones (Gibson & Shrader 2014). This could allow researchers to estimate the impact of a notification on engagement by comparing similar populations on the immediate side of time-zone borders over the course of an hour. Researchers should verify that the queue order in their context is not associated with potential confounders before using it as an instrument, as queues and the timing of notifications can be optimized to improve engagement.

Another engineering design where natural exogenous variation might be found at scale is in parallelized workloads, where data and tasks are split into smaller 'batches' and dispatched across a large number of nodes in a compute network. In some circumstances, batch membership might be found to be a valid instrument, as the compute time associated with certain tasks can vary widely based on a number of exogenous factors (network throughput, memory utilization, disk utilization, attributes of prior and concurrent tasks in a given data center, etc.).

These engineering designs and their implications for discovering exogenous variation at scale suggest the possibility of new types of collaborations between software engineers and empiricists in the system design phase of software development in order to maximize the information value of the system. For instance, perhaps notification systems could be designed to generate valid counterfactuals for estimating the impact of different notifications without requiring the overhead of an experimentation platform, while delivering similar levels of performance. This implies a different collaboration model than what we see at most firms, where experimentation platforms determine the ramp of new features and empiricists retrospectively recover, clean and analyze observational data on a feature's effectiveness for different subpopulations.

In addition to finding engineering designs where natural counterfactuals are likely to exist, automated approaches to finding natural experiments are another promising avenue of research for reducing search costs (Sharma et al. 2016).

Finally, we believe mobile and email notifications can serve as a valuable tool for implementing peer encouragement designs. Notifications are ubiquitous, and private to a specific user. Researchers can use notifications in combination with the network graph to generate causal estimates of peer effects in other domains.

**ACKNOWLEDGEMENTS**
We would like to thank Yiping Yuan, Derek Koh, Badrul Sarwar and Myunghwan Kim for their comments on the paper, and Yang Zhou and Ya Xu for their support and feedback on the project.

**REFERENCES**

Angrist, J. D., & Krueger, A. B. (1992). The effect of age at school entry on educational attainment: an application of instrumental variables with moments from two samples. Journal of the American statistical Association, 87(418), 328-336.

Angrist, J. D., & Krueger, A. B. (2001). Instrumental variables and the search for identification: From supply and demand to natural experiments. Journal of Economic perspectives, 15(4), 69-85.

Aral, S., & Walker, D. (2012). Identifying influential and susceptible members of social networks. Science, 1215842.

Aral, S., Muchnik, L., & Sundararajan, A. (2009). Distinguishing influence-based contagion from homophily-driven diffusion in dynamic networks. Proceedings of the National Academy of Sciences, 106(51), 21544-21549.

Asch, S. E. (1956). Studies of independence and conformity: I. A minority of one against a unanimous majority. Psychological monographs: General and applied, 70(9).

Bollinger, B., & Gillingham, K. (2012). Peer effects in the diffusion of solar photovoltaic panels. Marketing Science, 31(6), 900-912.

Coviello, L., Sohn, Y., Kramer, A. D., Marlow, C., Franceschetti, M., Christakis, N. A., & Fowler, J. H. (2014). Detecting emotional contagion in massive social networks. PloS one, 9(3), e90315.

Eckles, D., & Bakshy, E. (2017). Bias and high-dimensional adjustment in observational studies of peer effects. arXiv preprint arXiv:1706.04692.

Eckles, D., Kizilcec, R. F., & Bakshy, E. (2016). Estimating peer effects in networks with peer encouragement designs. Proceedings of the National Academy of Sciences, 113(27), 7316-7322.

Fiske, S. T., & Hauser, R. M. (2014). Protecting human research participants in the age of big data. Proceedings of the National Academy of Sciences, 111(38), 13675-13676.

Gibson, M., & Shrader, J. (2014). Time use and productivity: The wage returns to sleep. UC San Diego Department of Economics Working Paper.

Hartmann, W.R., Manchanda, P., Nair, H., Bothner, M., Dodds, P., Godes, D., Hosanagar, K. and Tucker, C., (2008). Modeling social interactions: Identification, empirical methods and policy implications. Marketing letters, 19(3-4), pp.287-304.

Kramer, A. D., Guillory, J. E., & Hancock, J. T. (2014). Experimental evidence of massive-scale emotional contagion through social networks. Proceedings of the National Academy of Sciences, 111(24), 8788-8790.

Manski, C. F. (2000). Economic analysis of social interactions. Journal of economic perspectives, 14(3), 115-136.

Nair, H. S., Manchanda, P., & Bhatia, T. (2010). Asymmetric social interactions in physician prescription behavior: The role of opinion leaders. Journal of Marketing Research, 47(5), 883-895.

Nam, S., Manchanda, P., & Chintagunta, P. K. (2010). The effect of signal quality and contiguous word of mouth on customer acquisition for a video-on-demand service. Marketing Science, 29(4), 690-700.

Nitzan, I., & Libai, B. (2011). Social effects on customer retention. Journal of Marketing, 75(6), 24-38.

Phan, T. Q., & Airoldi, E. M. (2015). A natural experiment of social network formation and dynamics. Proceedings of the National Academy of Sciences, 201404770.

Salganik, M. J., Dodds, P. S., & Watts, D. J. (2006). Experimental study of inequality and unpredictability in an artificial cultural market. science, 311(5762), 854-856.

Salganik, M. J. (2017). Bit by bit: social research in the digital age. Princeton University Press.

Saveski, M., Pouget-Abadie, J., Saint-Jacques, G., Duan, W., Ghosh, S., Xu, Y., & Airoldi, E. M. (2017, August). Detecting network effects: Randomizing over randomized experiments. In Proceedings of the 23rd ACM SIGKDD International Conference on Knowledge Discovery and Data Mining (pp. 1027-1035). ACM.

Shalizi, C. R., & Thomas, A. C. (2011). Homophily and contagion are generically confounded in observational social network studies. Sociological methods & research, 40(2), 211-239.

Sharma, A., Hofman, J. M., & Watts, D. J. (2016). Split-door criterion for causal identification: Automatic search for natural experiments. arXiv preprint arXiv:1611.09414.

Shi, C. & Fuad, A. Air Traffic Controller: Member-First Notifications at LinkedIn [Blog Post]. (2018, March). Retrieved from: https://engineering.linkedin.com/blog/2018/03/air-traffic-controller--member-first-notifications-at-linkedin

Ugander, J., Karrer, B., Backstrom, L., & Kleinberg, J. (2013, August). Graph cluster randomization: Network exposure to multiple universes. In Proceedings of the 19th ACM SIGKDD international conference on Knowledge discovery and data mining (pp. 329-337). ACM.

Yoganarasimhan, H. (2012). Impact of social network structure on content propagation: A study using YouTube data. Quantitative Marketing and Economics, 10(1), 111-150.